# *Yottixel* – An Image Search Engine for Large Archives of Histopathology Whole Slide Images

S. Kalra[a,b], C. Choi[b], S. Shah[b], L. Pantanowitz[c], H.R. Tizhoosh[a,d]

[a]*Kimia Lab, University of Waterloo, Ontario, Canada*
[b]*Huron Digital Pathology, St. Jacobs, ON, Canada*
[c]*University of Pittsburgh Medical Center, Department of Pathology, PA, USA*
[d]*Vector Institute, Toronto, Canada*

**Abstract**

With the emergence of digital pathology, searching for similar images in large archives has gained considerable attention. Image retrieval can provide pathologists with unprecedented access to the evidence embodied in already diagnosed and treated cases from the past. This paper proposes a search engine specialized for digital pathology, called *Yottixel*, a portmanteau for "*one yotta pixel*," alluding to the big-data nature of histopathology images. The most impressive characteristic of Yottixel is its ability to represent whole slide images (WSIs) in a compact manner. Yottixel can perform millions of searches in real-time with a high search accuracy and low storage profile. Yottixel uses an intelligent indexing algorithm capable of representing WSIs with a *mosaic* of patches by converting them into a small number of methodically extracted barcodes, called "Bunch of Barcodes" (BoB), the most prominent performance enabler of Yottixel. The performance of the prototype platform is qualitatively tested using 300 WSIs from the University of Pittsburgh Medical Center (UPMC) and 2,020 WSIs from The Cancer Genome Atlas Program (TCGA) provided by the National Cancer Institute. Both datasets amount to more than 4,000,000 patches of $1000\times1000$ pixels. We report three sets of experiments that show that Yottixel can accurately retrieve organs and malignancies, and its semantic ordering shows good agreement with the subjective evaluation of human observers.

## 1. Motivation

Large archives of digital scans in pathology are gradually becoming a reality. The amount of information stored in such archives is both impressive and overwhelming. However, there is no convenient provision to access this stored knowledge and available to pathologists for diagnostic, research, and educational purposes. Traditionally, a medical image database (including a digital pathology archive) stores images tagged with some metadata mainly in the form of textual information (i.e., reports). A medical practitioner (e.g., a pathologist) can query his/her own medical search terms in such databases to find the images associated with those keywords. However, a text-based search query is incapable of exploiting the intrinsic nature of anatomical and morphological clues in medical images, thus delivering very limited information access to end users. In fact, text descriptions are generally inadequate to comprehensively index the content of medical images. This is sometimes because of the sheer amount of effort required to annotate a large number of images. However, more importantly, it is simply not possible to describe the pervasive variability of complex anatomical patterns in medical images with nearly sufficient distinction for retrieval purposes in most cases. For example, a single pathology image may contain just basic types of tissues (e.g., epithelium and connective tissue). However, the actual number of visual patterns derived from these basic tissues, from a computer-vision perspective, is nearly infinite (Tizhoosh and Pantanowitz; Gurcan et al., 2009b; Niazi et al., 2019). Apparently, domains such as medical image analysis are not profiting much from text-based image search as much as other applications because the most helpful search tasks have to address the visual manifestation of malignancies in the query tissue (Tizhoosh et al., 2016; Caicedo et al.). These challenges necessitate inquiry into more effective ways of searching in medical image archives. Digital pathology, because of the image size, complexity and color, and definitiveness of diagnosis at the pathology level, has attracted a considerable amount of attention. The general approach for searching in a medical image database is content-based image retrieval (CBIR) (Figure 1). To facilitate image search, CBIR algorithms essentially describe the content of an image with non-textual attributes, generally with a vector of real numbers known as a *feature vector*. If a feature vector encompasses the descriptive visual properties of an image, then searching for similar images becomes a nearest-neighbour matching problem. Images with similar content could be retrieved based on a comparison of their feature vectors and not based on the associated textual metadata. This is generally possible if a feature vector encodes the semantic structures of an image invariant to scale, rotation, translation, and even to some degree, to deformation. Such rich and descriptive features can *characterize* images for the purpose of identification, which is the core task of any CBIR system.

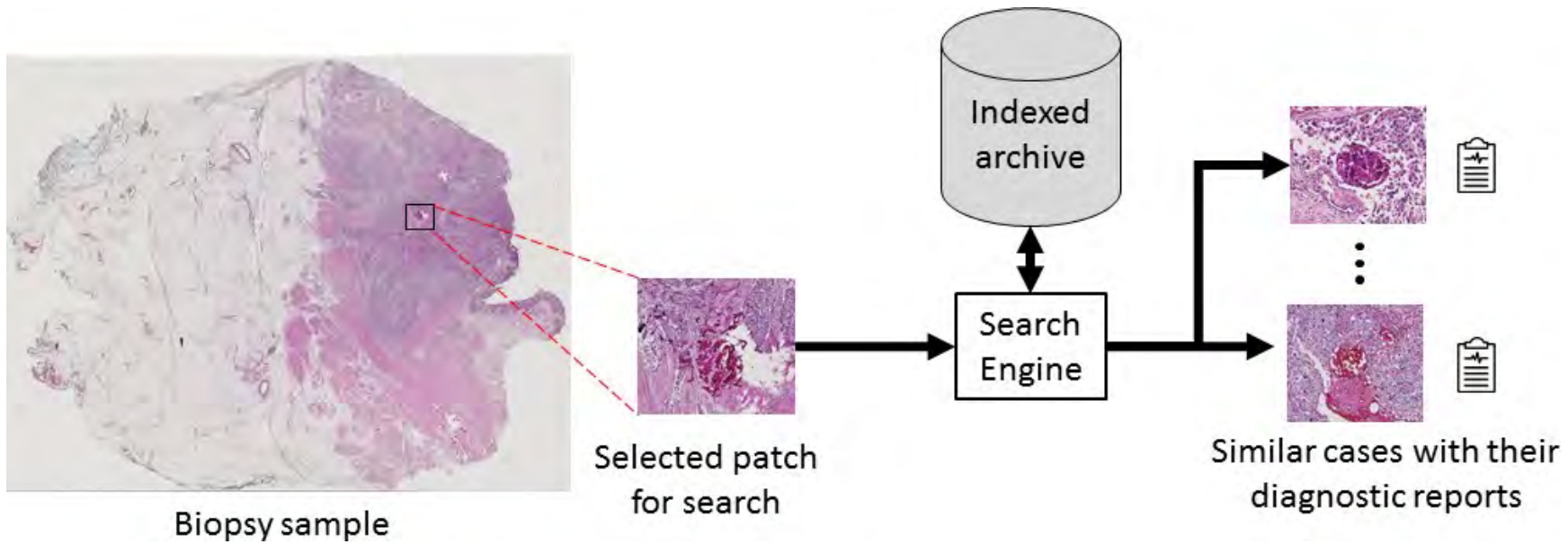

Figure 1: General workflow of CBIR systems for digital pathology.

CBIR systems customized for histopathology can exploit evidence-based knowledge from past cases and make them available to pathologists for more efficient and more informed decision making. However, there are two major drawbacks of CBIR systems that limit their integration into digital pathology. Firstly, most CBIR proposals use basic image features that capture low-level characteristics of an image such as color, edges, textures, or shapes. This approach generally fails to capture high-level patterns corresponding to the semantic content of histopathology images. Secondly, whole slice images (WSIs) deal with gigapixel digital images of extremely large dimensions (i.e., larger than 50000 $\times$ 50000 pixels). However, most proposed CBIR technologies are designed for natural images that have smaller dimensions (i.e., smaller than 300 $\times$ 300 pixels). In addition to the large dimensions, pathology images exhibit an intractable level of variability in visual features that makes their identification, compared with that of natural images, even more challenging. For instance, histology and histopathology images contain several diversely shaped edges, intricate and irregular structures, and high gradient changes that create an inconceivable complexity for most computer vision algorithms. Wehave designed and developed a Yottixel prototype with these limitations in mind. We are training/using deep architectures topologies capable of identifying semantically related images rather than just visually similar images, for feature extraction. This is accompanied by an intelligent indexing technique for representing a WSI in a compact and memory-efficient way by using a “bunch of barcodes.” These strategies and heuristics provide better search results, which are more likely to match image content in a manner pathologists would desire.

This paper is organized as follows: section 2 presents a review of previous works related to histopathology image retrieval. section 3 describes the algorithms behind Yottixel. section 4 describes the two datasets used in this study. The experimental setup and results are presented in section 5.

## 2. Related Works

Contrary to popular belief, digital image analysis was not adopted for face recognition. Rather, it was used for the study of medical images (Madabhushi and Lee, 2016). A survey article (Gurcan et al., 2009a) suggests the widespread use of computer assisted diagnosis (CAD) can be traced back to the development of digital mammography during the early 1990s. In fact, CAD is now integral to many clinical routines for diagnostic radiology and recently becoming eminent in diagnostic pathology as well.

With an increase in the workload of pathologists, there is a compelling need to integrate CAD systems into pathology routines (Komura and Ishikawa, 2018; Madabhushi and Lee, 2016; Madabhushi et al., 2011; Gurcan et al., 2009a). Researchers in both image analysis and pathology fields have recognized the importance of the quantitative analysis of pathology images by using machine learning (ML) techniques (Gurcan et al., 2009a). The continuous advancement of digital pathology scanners and their proliferation in clinics and laboratories has resulted in a substantial accumulation of histopathology images, justifying the increased demand for their analysis to improve the current state of diagnostic pathology (Madabhushi and Lee, 2016; Komura and Ishikawa, 2018).

**Image retrieval**, which implies searching for similar images, requires extracting salient features that are descriptive of image content. In its entirety, there are two main methods for processing WSIs (Barker et al., 2016). The first method is called *sub-setting method*, which considers a small section of a large pathology image as an essential part, such that the processing of a small subset substantially reduces processing time. A majority of research studies in the literature have used the *sub-setting* method because of its speed and accuracy. However, it requires expert knowledge and intervention to extract the proper subset. On the other hand, the *tiling method* segments images into smaller and controllable patches (i.e., tiles) and tries to process them against each other (Gutman et al., 2013), which will naturally require a careful approach toward design and will be more expensive. However, it indeed is a distinct approach toward full automation.

Traditionally, large medical image archives contain textual annotations to facilitate search; however, the performance of this approach is not good to locate anatomical similarities. In 2003, an online CBIR system was developed wherein the client provides a query image and the corresponding search parameters to the server side (Zheng et al., 2003). The server then performs similarity searches based on the feature types, such as color histogram, image texture, Fourier coefficients, and wavelet coefficients, while using vector dot product as a distance metric for retrieval. The server then returns images that are similar to the query image along with similarity scores and feature descriptor.

On the other hand, other works proposed an offline CBIR systems that utilizes sub-images rather than the entire digital slide (Mehta et al., 2009). Scale-invariant feature transform (SIFT) was used to index each sub-image for searching similar structures (Lowe, 1999). The experimental results suggested that 80% accuracy for the top 5 results retrieved from the database that holds 50 IHC stained pathology images (immunohistochemistry), consisting of 8 resolution levels, compared with manual search. Researchers have also developed a multi-tiered CBIR system based on WSI, which is capable of classifying and retrieving digital slides by using both multi-image query and images at the slide level (Akakin and Gurcan, 2012). The authors tested the proposed system on $1,666$ WSIs extracted from 57 follicular lymphoma (FL) tissue slides containing three subtypes and 44 neuroblastoma (NB) tissue slides comprising 4 sub-types. Experimental results suggested 93% and 86% average classification accuracy for FL and NB diseases, respectively.

More recently, a scalable CBIR method has been developed to cope with WSI by using the supervised kernel hashing technique that compresses a 10,000-dimensional feature vector into ten binary bits, which is reportedly observed to be a suitable representation of the image (Zhang et al., 2015). These short binary codes are then used to index all existing images for rapidly retrieving new query images. The proposed framework is validated on breast histopathology data set comprising 3,121 WSIs from 116 patients. The experimental results indicate an accuracy of 88.1% for processing at a speed of 10 ms for all 800 testing images.

Recently, hashing methods have been intensively investigated in the ML and computer vision community for large-scale image retrieval. Representative methods include, but are not limited to, weakly supervised hashing in kernel space (Mu et al., 2010), semi-supervised hashing (Wang et al., 2010), supervised hashing (Liu et al., 2012a), and compact kernel hashing with multiple features (Liu et al., 2012b). Among these methods, kernelized and supervised hashing (KSH) (Liu et al., 2012a) is generally considered the most effective, achieving state-of-the-art performance at a moderate training cost. The central idea of KSH is to reduce the gap between low-level hash code similarity and high-level semantic (label) similarity by virtue of supervised training. In doing so, a similarity search in the binary code space can reveal the given semantics of examples. In other words, KSH does well in incorporating the given semantics into the learned hash functions or codes, while the other hashing methods cannot leverage the semantics adequately. Specifically, KSH has a higher search accuracy compared with those of the unsupervised kernel hashing method and the semi-supervised linear hashing method, as it takes full advantage of supervised information (originating from the semantics) that is not well exploited by unsupervised and semi-supervised methods. KSH still shows clear accuracy gains at a much shorter training time compared with those of competing supervised hashing methods such as binary reconstructive embedding (BRE) (Kulis and Darrell, 2009) and minimal loss hashing (MLH) (Norouzi and Blei, 2011). One major limitation of KSH is that the optimization required to obtain good hash functions is time consuming.

The motivation for developing *Yottixel* emerged from multiple observations: 1) Not many works have provided a search solution for WSIs; the focus is generally on patch processing (for instance, (Galaro et al., 2011; Sharma et al., 2012; Vanegas et al., 2014)), 2) Much research has been dedicated to process *labelled* repositories where malignant regions in WSI files have been delineated by trained pathologists (for instance, (Wang et al., 2016; Barker et al., 2016; Liu et al., 2017)) 3) Many approaches index images with real-valued features, a requirement that would be hard to meet in reality because of storage and computational requirements (Wan et al., 2014; Yang et al., 2013; Ma et al., 2017) 4) Some works use hashing for fast search to increase the feasibility of retrieval, but hash codes may not easily facilitate data exchange among repositories (for instance, (Jiang et al., 2016; Yang et al., 2018; Shi et al., 2017; Do et al., 2019)).

The design of Yottixel is meant to provide a search engine that processes unlabelled images through explicit binary codes for fast search and easily transferable index between different locations. In addition to using deep features, Yottixel makes this possible through three key ideas: 1) color clustering in low magnification, 2) assembling a "mosaic" of representative patches encoded in deep features, and 3) the crucial concept of "bunch of barcodes".

## 3. Yottixel – A Search Engine for WSI

This section describes the design and implementation of the proposed image retrieval and index framework, Yottixel (Figure 2). This work has two main practical contributions. First, we propose a method for representing an entire WSI with a small set of patches, referred to as *mosaic* (Fig. 3). The concept of mosaicking is fundamental for the feasibility of image search. Secondly, we construct and test an end-to-end ensemble framework that indexes and retrieves WSIs based on their content at a high speed and requires low storage.

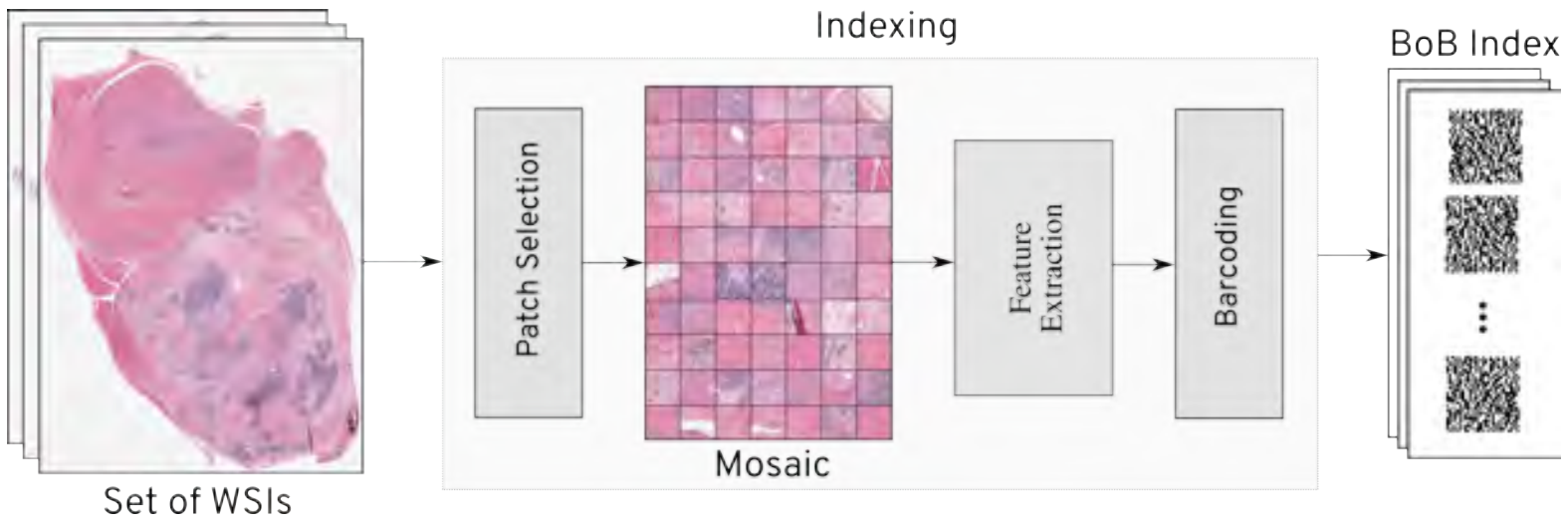

Figure 2: Overview of Yottixel's indexing framework to generate the BoB index. Patch selection (Fig. 3) generates the mosaic. Individual barcodes may be used for patch search. All barcodes of any given scan can be used for searching WSI.

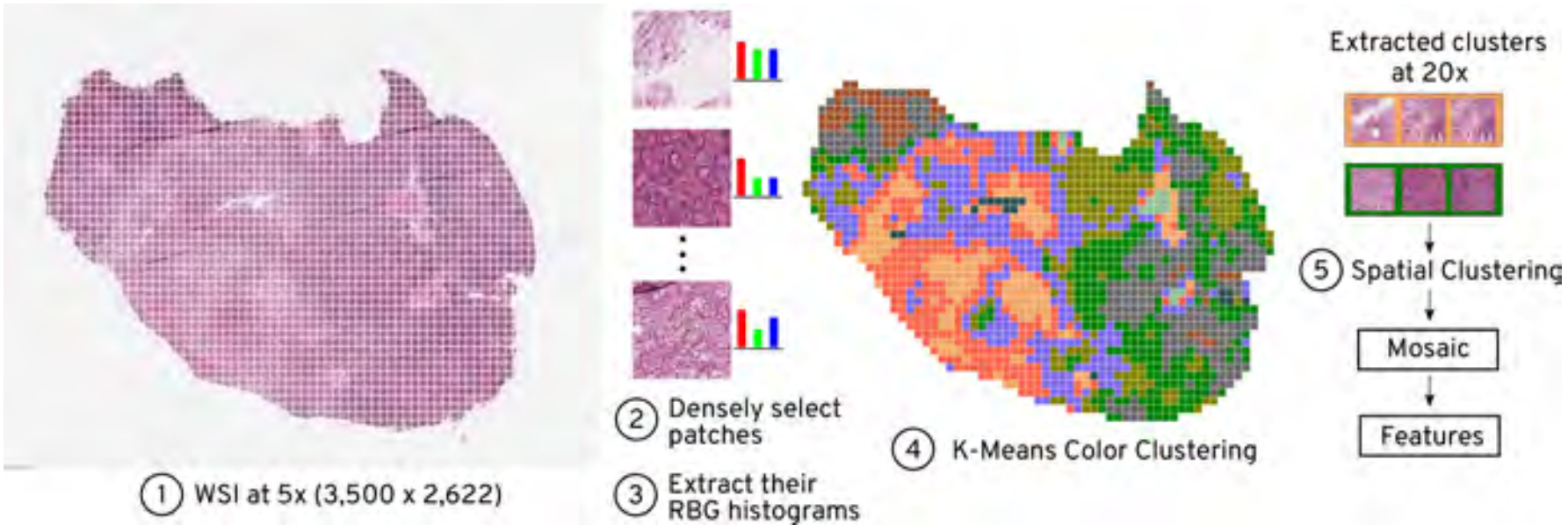

Figure 3: A schematic of the algorithm for creating mosaic from a WSI. The concept of *mosaicing* is paramount for real-time and efficient search in large archives of WSI files. A WSI is split into a dense grid of *patches*(or tiles) represented by some features (e.g., RGB histograms). Clustering the features, through algorithms such as k-means, can build the basis for assembling a mosaic of patches that represent the entire WSI.

The distinguishing aspect of Yottixel is the utilization of barcodes for image representation and characterization. A WSI is indexed by converting its associated mosaic to a set of barcodes. This set of barcodes constitutes an index for the given WSI, referred to as "**Bunch of Barcodes**" (BoB) index. The BoB index accelerates the retrieval process and alleviates the computation and storage burden on the deployment infrastructure for laboratories and clinics. Yottixel is a complete and functioning search engine for indexing WSIs for CBIR systems with major emphasis on performance and scaling for laboratory and hospital requirements.

Yottixel has two major phases of operation: *(i)* offline indexing and *(ii)* run-time search. During the initial deployment of Yottixel, offline indexing consumes the maximum computation resources to index the available WSI files (hence, we strongly recommend implementing Yottixel into scanners to index images when they are captured using the GPU power of the scanner). Once a sufficient number of images are indexed, the two phases are activated simultaneously. However, offline indexing is set to run preemptively allowing runtime search to acquire higher precedence over the available resources.

### *3.1. Offline Indexing Phase*

The crux of a search engine platform for a large archive of medical images of high dimensionality is its **indexing**. The structure of the index determines the speed, reliability, and robustness of search results. Yottixel indexes a WSI by *(i)* computing its mosaic (a representative set of patches) and then *(ii)* converting the mosaic to a BoB index. The design choices of the indexing algorithm are influenced by real-world scenarios in a mid-size pathology laboratory or clinic, where hundreds of thousands of WSI files are generated every year. However, the computing and storage infrastructure are generally not sufficient for hosting a sophisticated image search engine on-site. Indexing cannot be implemented in hospitals and laboratories because of the requirement of high storage and computational resources amid a sluggish transition to digital pathology that requires an expensive IT infrastructure.

**Computing the Mosaic**: Yottixel receives a set of WSIs queued for indexing. For each queued WSI, a representative set of patches, or *mosaic* is computed (Fig. 3). Employing a mosaic considerably reduces the computation burden. Instead of operating over an entire WSI, all the subsequent image processing operations are applied on the mosaic of WSI. The algorithm for creating a mosaic is outlined in Algorithm 1, Lines 8–26. Firstly, a WSI is segmented into nine different regions based on their colour composition by using the k-means algorithm. The number of tissue regions is a parameter decided by manually inspecting different WSIs. We found that a typical WSI exhibits a maximum of nine different types of visually distinct regions. Segmenting these regions captures the variability from a computer vision perspective but may not have relevance strictly from a histopathology point-of-view. However, colour-based segmentation frequently resulted in the separation of different tissue types within a WSI, such as blood stains from muscles, fat, and in some cases, even cancerous regions. From these segmented regions, a small percentage of patches are randomly selected (e.g., 5%) while preserving the spatial diversity, again using the k-means algorithm (Algorithm 1, Lines 21–26). The patches are collected from all segmented regions constituting the mosaic of the given WSI (Fig. 4). Patch clustering may be performed at a lower resolution (e.g., $5\times$ magnification) because a higher resolution does not offer any superiority. With these settings, a typical mosaic obtained is $\approx 20$ times smaller than the specimen area depicted in the WSI.

**Creating the BoB Index**: The patches in a mosaic are converted to a set of barcodes. This BoB constitute the index for a single WSI file. The algorithm for creating the BoB index from a given WSI is provided in Algorithm 1, lines 27–35. First, a patch is converted to a feature vector using a deep network (we used trained, pre-trained, and fine-tuned deep networks). Although pre-trained networks have learned from natural images, they may still offer robust image characterization properties for histopathology images. We mostly experimented with the last average pooling layer of the VGG19 (Simonyan and Zisserman, 2014), Inception (Szegedy et al., 2015), Densenet (Huang et al., 2017), and in-house trained and fine-tuned solutions (Babaie et al., 2017; Kieffer et al., 2017) to extract feature vectors from mosaic patches. Then, we used the discrete differentiation (or MinMax algorithm (Tizhoosh et al., 2016), Fig. 5) to convert the feature vector to a binary representation called "barcode." The BoB index is light-weight and enables a fast Hamming distance search. For an average WSI file of size $\approx 700$ MB, the BoB index can be as small as $\approx 10$ KB, i.e., $70,000$ smaller than the original file.

**Binarization using MinMax algorithm:** Although deep features can be used directly to measure the similarity between images via distance metrics such as $L_2$, computational efficiency is a serious issue, especially for searches in large databases across all primary sites (i.e., exhaustively searching k-nearest neighbors). Therefore, we employed a binarization method to convert these features into binary codes. Binary features allow for fast real-time search. During a run-time query, high-dimensional features are extracted from the query image and converted to barcodes. We used accelerated CPU commands to calculate the Hamming distance for the nearest neighbors queries. It has been stated that the MinMax algorithm for binarization is particularly useful for the retrieval and indexing of histopathology scans in terms of both speed and storage (Kumar et al., 2018). Furthermore, empirical evidence from our experiments validates the latter claim, suggesting that the technique is simple yet effective.

### *3.2. Runtime Search Phase*

Once a sufficiently large index is created, Yottixel provides users with an interactive interface to perform search queries on their WSIs. There are two modes of searching—vertical and horizontal. In the **vertical search** mode, image matching is confined to the same primary site as the query patch, whereas in the **horizontal search**, the entire index is searched across all primary sites.

## 4. Datasets

To validate the search capabilities of *Yottixel*, we utilized two different datasets. The first is a private dataset consisting of 300 H&E stained WSIs across more than 80 different primary diagnoses from multiple organs provided by the University of Pittsburgh Medical Center (UPMC). The total size of the dataset in the compressed form was 104 GB. The second dataset consists of 2,020 WSIs taken from a public repository of more than 33,000 WSIs from the NIH's pan-cancer analysis project *The Cancer Genome Atlas* (TCGA) (Weinstein et al., 2013). The total size of this dataset in the compressed form is 2 TB. An average WSI within both the datasets is $\approx 45{,}000 \times 45{,}000$ pixels. Each WSI in both datasets was labeled with the type of malignancy (primary diagnosis) and the affected organ (primary site) as text values. For the UPMC dataset, these text values are not normalized and contain many redundancies. We performed various pre-processing tasks on the WSIs as follows:

**WSI Removals** – A WSI was removed from the dataset if one or more of the following was present: poor staining, low resolution, only one available level in WSI's pyramid, and/or presence of large out-of-focus regions. In total, five slides were removed from the UPMC dataset and seven slides from the NIH dataset.

**Preprocessing** – The tissue regions within the slides were segmented using common thresholding methods and manual segmentation. Two such examples are shown in Fig. 6. For some cases in the UPMC dataset, automatic segmentation did not perform well, and we corrected them manually. For the larger NIH dataset, no manual intervention was performed.

**Class Imbalance** – There is a high class imbalance in both datasets, which is quite common in histopathology archives (mainly because of the varying incidence rates of different malignancies). In the UPMC dataset, out of $\approx 80$ classes, 30% of the WSIs belong to the top four classes–lung, skin, soft tissue, and brain. Approximately 60% of WSIs in both datasets are distributed across the first quartile of the categories. The distribution of the WSIs for the UPMC dataset across the top 15 primary sites and diagnosis are listed in Table 1. A similar trend is observed for the NIH dataset (Table 2, Fig. 7). The primary diagnosis associated with slides are not pre-processed and contains redundant values. For instance, there is no defined hierarchy among "Adenocarcinoma" or "Prostatic Adenocarcinoma." Basic preprocessing such as spelling errors and text normalization were applied. The search algorithm of Yottixel does not utilize these labels for training. We only use them for validating search results.

| Primary Diagnosis | Count |
|---|---|
| Prostatic adenocarcinoma | 5 |
| Merkel cell carcinoma | 4 |
| Adenocarcinoma | 4 |
| Pleomorphic adenoma | 4 |
| Adenoid cystic carcinoma | 3 |
| Differentiated liposarcoma | 3 |
| Papillary thyroid carcinoma | 3 |
| Neurofibroma | 3 |
| Granular cell tumor | 3 |
| Metastatic lobular breast carcinoma | 2 |
| Basal cell adenocarcinoma | 2 |
| Squamous cell carcinoma | 2 |
| Invasive moderately-Differentiated adenocarcinoma | 2 |
| Canalicular adenoma | 2 |
| Metastatic seminoma | 2 |

| Primary Site | Count |
|---|---|
| Lung | 25 |
| Skin | 23 |
| Soft tissue | 21 |
| Brain | 15 |
| Prostate | 15 |
| Salivary gland | 13 |
| Kidney | 11 |
| Thyroid | 10 |
| Breast | 9 |
| Adrenal | 7 |
| Lymph node | 7 |
| Thoracic | 6 |
| Testis | 6 |
| Colon | 5 |
| Bone | 5 |

Table 1: Number of WSI files in the UPMC dataset across top 15 primary sites and diagnoses. Class imbalance is evident specially across the primary sites with a majority of files belonging to lung, skin, and soft tissue samples.

| Primary Diagnosis | Count |
|---|---|
| Breast Invasive Carcinoma | 189 |
| Glioblastoma Multiforme | 165 |
| Brain Lower Grade Glioma | 164 |
| Uterine Corpus Endometrial Carcinoma | 123 |
| Kidney Renal Clear Cell Carcinoma | 116 |
| Skin Cutaneous Melanoma | 101 |
| Sarcoma | 100 |
| Colon Adenocarcinoma | 99 |
| Head and Neck Squamous Cell Carcinoma | 98 |
| Lung Squamous Cell Carcinoma | 95 |
| Lung Adenocarcinoma | 94 |
| Thyroid Carcinoma | 89 |
| Prostate Adenocarcinoma | 79 |
| Stomach Adenocarcinoma | 64 |
| Liver Hepatocellular Carcinoma | 61 |
| Cervical Squamous Cell Carcinoma and Endocervic... | 51 |
| Kidney Renal Papillary Cell Carcinoma | 50 |
| Bladder Urothelial Carcinoma | 40 |
| Pheochromocytoma and Paraganglioma | 37 |
| Esophageal Carcinoma | 31 |
| Testicular Germ Cell Tumors | 28 |
| Adrenocortical Carcinoma | 21 |
| Ovarian Serous Cystadenocarcinoma | 19 |
| Rectum Adenocarcinoma | 19 |
| Pancreatic Adenocarcinoma | 17 |
| Uveal Melanoma | 14 |
| Kidney Chromophobe | 14 |
| Mesothelioma | 12 |
| Lymphoid Neoplasm Diffuse Large B-cell Lymphoma | 10 |
| Uterine Carcinosarcoma | 10 |
| Thymoma | 9 |

Table 2: Number of WSI files in the NIH dataset (subset of TCGA) across different primary diagnoses.

## 5. Experiments and Results

This section reports the results of several experiments performed using the two datasets described in the previous section. We performed all experiments on a Dell EdgeServer Ra with 2x Intel(R) Xeon(R) Gold 5118 (12 cores, 2.30GHz), 2x Telsa V100 (v-RAM 32 GB each), and 394 GB RAM. The code for indexing is written in C/C++, whereas the UI components are written in multiple languages, but mostly in Python and Javascript.

**Parameters:** A set of parameters for indexing was set empirically. The number of color clusters $k_{CH}$ was set to 9. The percentage of patches $p_M$ to build the mosaic was set to 5%. Clustering was performed in $m^c_x = 5x$ whereas indexing was performed in $m^{idx}_x = 20x$. The patch size at low magnification was $s_l = 250$ pixels (equivalent to $2mm$) and $s_h = 1000$ pixels (equivalent to $500\mu m$).

**Deep features:** With the exception of the VGG network, all other networks (Inception, DenseNet, and in-house trained CNNs) did in fact provide comparable results. All reported experiments used DenseNet features.

*5.1. Experiment Series 1: Scan-to-Scan Matching*

In the first experiment series, the performance of the scan search was evaluated. Given a query WSI $I_q$, we retrieve a set of similar WSIs $R$ not containing $I_q$, i.e. $R \cap I_q = \varphi$. We define a single experiment as running the retrieval task for a set of WSIs that have similar attributes either the same primary site or diagnosis. A retrieval task is considered successful if there is at least one WSI within the retrieved set $R$ that has the same attribute (primary site or primary diagnosis) as that of the query WSI $I_q$. The accuracy is defined as the average leave-one-out accuracy of the retrieval experiments for any given configuration of the attribute.

The algorithm to determine the distance between two given WSIs is outlined in Algorithm 2. It is not commutative, which means that the order of arguments would affect the outcome, or the distance between $A$ and $B$ is different from that $B$ and $A$. The intuition behind computing the distance between two WSIs is based upon the Hamming distances among the barcodes of their BoB indexes. More specifically, the distance of a query WSI $I_q$ from another WSI $I$ is the median of the minimum Hamming distances obtained from each barcode in the BoB of $I_q$ to the BoB of $I$. The best match, hence, is the one with the median value of the minimum Hamming distance among the patches of two mosaics.

We performed the experiments using WSIs from four different types of primary sites and diagnoses, retrieving the **top ten** WSIs in each experiment. We chose the lung, brain, breast, and thymus for the primary site, and lung adenocarcinoma, brain lower grade glioma, kidney chromophobe, and rectum adenocarcinoma for primary diagnosis. We chose these cases because they were more frequent in the dataset (hence, delivering more statistics for experimenting) and also offered high variability (hence, sufficiently challenging). We used accuracy and number of correct retrievals as the validation metrics for the experiments.

**Impact of the size of mosaic** – We randomly selected four different portions of the indexed mosaic (10%, 30%, 70%, 100%) for each experiment. Because we randomly selected the mosaic subset, we performed each experiment 50 times (except when we selected the entire mosaic). For each experiment, we recorded the mean and standard deviation of matching hits. Therefore, the total number of experiments was $10 \times 4 \times 8 = 320$, out of which 240 experiments were run 50 times each ($10 \times 8 = 80$ experiments were run only once, whereas $320 - 80 = 240$ experiments were run 50 times each).

The accuracy values obtained from all the experiments are shown in Fig. 10. The red dotted curves/lines in the graphs show the probability of success if the WSIs were to be selected randomly from the dataset. Except for the rectum adenocarcinoma, all other graphs show that accuracy values reach ~100% within ten retrievals. In other words, if we retrieve the top ten WSIs, at least one WSI can be found among the retrieved WSIs with the same attribute as the query WSI. The results show that our approach performs three to four times better than the random approach. However, for the retrieval of thymus, or WSIs with kidney chromophobe and rectum adenocarcinoma, the performance of our search retrieval method was considerably better than the random approach. We expect that the margin should be much higher for larger datasets. For most experiments, a smaller mosaic subset has no significant performance improvement over the larger mosaic (except for the thymus). This finding suggests that we can further reduce the mosaic size for scan-to-scan retrieval. However, the specificity for patch retrieval is lost as the number of patches in the mosaic will not be sufficient for the retrieval.

From the experiments, we can conclude that a smaller sized mosaic can be as accurate as a larger mosaic for the scan-to-scan matching, especially when the retrieval task is to match the primary diagnosis. We keep higher number of patches in our mosaic because it aligns with the premise on which our search engine is based—to index WSIs in their entirety and give freedom to pathologists to perform the localized searches. We hypothesize that our platform will mainly be used to make localized searches (patch-to-patch) instead of searching the entire WSI (scan-to-scan), therefore having a greater number of patches in the index would improve the specificity of the search results. Keeping all the patches from a WSI would be the ideal case; however, this would require considerable storage capacity. Therefore, we have set a "sweet spot" of 5% patches to be contained in our mosaics.

**Analysis of number of correct retrievals –** For the next series of experiments, we fixed the mosaic size to be 100%, of what has been indexed which is 5% of number patches in a scan). Then, we measure the number of correct retrievals in each experiment. The results are summarized in Fig. 11. The red dotted lines/curves represent the expected number of correct retrievals if each retrieval task is considered as a Bernoulli trial. Note that our retrieval tasks cannot be accurately modeled as Bernoulli trials, but they are easy approximations. The real expected value should be even lower than the red dotted lines/curves. From the graphs, one can clearly see that Yottixel is far superior to the random retrieval. The large discrepancy between the random retrieval and the Yottixel search validates that our search engine can capture the semantics of a WSI pertaining to its malignancy or general anatomy. The WSIs from the brain are the least confusing to our search engine. The lung and breast have almost the same random retrieval performance (indicated by the red lines) as that of the brain. This could be attributed to the polymorphic nature of the WSIs from the lung and breast.

### *5.2. Experiment Series 2: Classification*

In the second experiment series, we measure the performance of our search engine based on its classification capability. It is important to emphasize that the Yottixel is not primarily designed for classification. However, it is a reasonable assumption that a good search engine should draw a distinct decision boundary within images from different semantic categories. We performed vertical searches (limited within the same primary site) to determine the accuracy of the Yottixel in distinguishing among different tumor types. We used a *majority voting* among the top five search results to assign a "class" (i.e., primary diagnosis) to a given query WSI. The performance of the search engine as a classification model is summarized as confusion matrices in Fig. 12. We chose six different primary sites: 1) adrenal gland, 2) brain, 3) kidney, 4) colorectal, 5) uterus, and 6) lung. These primary sites were chosen because they offered WSIs with more than two types of primary diagnosis. The aim of our search engine is to distinguish between these different primary diagnoses within the same primary site. The lowest accuracy of 69.84% is obtained for the lung —lung adenocarcinoma and lung squamous Cell carcinoma. The highest accuracy of 93.20% is obtained for the adrenal gland uterus—uterine carcinosarcoma and uterine corpus endometrial carcinoma. We also obtained >90% accuracy for adenal gland distinguishing adrenocortical carcinoma and pheochromocytoma & paraganglioma.

Note that treating the search as a classification may be too restrictive: two differently diagnosed tissue samples may still be reasonably similar with some diagnostic value. Throughout the body there are tumors and lesions that overlap morphologically. As a result, many of these entities have similar patterns (Choi and Ro, 2018). Hence, the pattern often needs to be combined with ancillary studies including immunohistochemistry and molecular testing for an accurate diagnosis.

### *5.3. Experiment Series 3: Testing by Users*

Generally, because no labeled images are produced during the clinical workflow in digital pathology, in the third experiment series, we measured the accuracy of search and retrieval through user feedbacks. We evaluated how well our search results align with the subjective perception of its users. The Yottixel search results were evaluated by an expert user (a pathologist, co-author LP) and six non-clinical users with computer vision experience (Table 4, Fig. 13).

For this experiment series, we created a web application to gather the user's subjective evaluations about the search results. The web application presented a user with query images and top three search results in a random order. The users were not aware of the order. At the same time, the users were not aware that the search results are the top three images. In each session, there were a total of 48 queries. All participants answered the same questions, but we reshuffled the questions and ordering in each session to counteract any biases.

For each query result, participants would provide their feedback from five discrete values ranging from `Bad` (red) to `Great` (green). After gathering the data for the study, we sorted the participant feedback in the original order per search engine, referring $Q_1$, $Q_2$, and $Q_3$ as the top three results, respectively. In our analysis, we treated the expert user feedback (pathologist, LP) differently from non-expert users.

The general summary of the participant's feedback is presented in Fig. 13. Both expert and non-expert users, ranked $Q_1$ more positively than $Q_3$. It is interesting to note that, on an average, non-expert users ranked a higher number of `Very Bad` to $Q_1$ compared with $Q_2$ and $Q_3$. However, this was not true for the pathologist. The trends for the pathologist are very concrete and reflect positively on our approach. For instance, $Q_1$ has the highest number of `Great` compared with others, $Q_2$ has higher number of `Great` and `Good` than $Q_3$. Similarly, $Q_3$ has the highest number of `Very Bad`.

The Good and Great refers to the agreement, whereas Bad and Very Bad refers to disagreement of the query result. We aggregated the responses into three broad categories: disagreement, neutral, and agreement. The results of the response aggregation are summarized in Table 4. Furthermore, Fig. 14 shows an interesting trend of the Hamming distance versus the selected option by the pathologist. It shows that the median Hamming distance of images marked as Great are much lower than the ones marked Very Bad. This further validates the efficacy of the underlying metric used by the Yottixel to calculate the similarity among patches.

## 6. Summary and Conclusions

In this work, we introduced Yottixel, an image search engine for digital pathology. Content-based image retrieval identifies digital images using pixel values and their features. Digital pathology can benefit from image search in large archives of WSIs as a dynamic and smart platform to exploit the information stored in evidently diagnosed cases. The image search is based on a combination of supervised and unsupervised algorithms. Deep features are employed to characterize images (i.e., patches). The search technology is inherently "unsupervised" because it works with raw data with no specific training for the search task. We use different algorithms including segmentation and clustering algorithms, deep networks, and distance metrics for search and retrieval. The proposed image search platform was tested with a private dataset of 300 WSIs and also with a public dataset of 2025 WSIs. The initial results of our validation experiments are quite encouraging. The visual similarity of the retrieved cases for most queries images were striking when evaluated by both expert and non-experts. The search experiments results were accurate when treated as a classification, and they exhibited overall good conformance with the expert's evaluations. Our preliminary results on a small but extremely diverse private dataset (i.e., UPMC) and a medium-sized public dataset, namely a portion of the TCGA archive (but still the largest reported thus far for image retrieval in pathology) both demonstrate the feasibility of the proposed technology and justify further investigations. We continue to improve the accuracy and speed requirements of the Yottixel platform to make it more usable for diagnostics, research and educational purposes.

**Acknowledgements –** The funds for this research have been provided by the ORF-RE program (Ontario Research Fund - Research Excellence). Core research was also supported by NSERC (Natural Sciences and Engineering Research Council of Canada). The first author's internship at Huron Digital Pathology is supported by MITACS (Mathematics of Information Technology and Complex Systems).

**Algorithm 1** Pseudo-code for creating the index or bunch of barcodes (BoB) for a given WSI $I$

Set $k_{CH}$ (number of color clusters)
Set $p_M$ (percentage of patches to build the mosaic)
Set $m_x^c$ (clustering magnification)
Set $m_x^{idx}$ (indexing magnification)
Set patch sizes $s_l/s_h$ in low/high magnifications
**procedure** Create_Index($I$)
▷ Extract the tissue regions
$T \leftarrow$ TissueSegmentation($I$)
▷ Select a low magnification within the WSI pyramid
$I_{m_x^c} \leftarrow$ SelectMagnification($I$, $m_x^c$)
    ▷ Perform dense patching for patch size $s_l \times s_l$
    $P \leftarrow$ DensePatching($I_{m_x^c}$, $s_l$)
    ▷ Isolate patches containing tissue regions
    $P_T \leftarrow T \cap P$
    **for** $i \in [1, Len(P_T)]$ **do**
        ▷ Calculate the histogram of $i^{th}$ patch
        $H_{P_T}[i, :] \leftarrow$ RGBHistogram($P_T[i]$)
    **end for**
    ▷ Perform k-means clustering on histograms
    $C_1, C_2, \ldots, C_{k_{CH}} \leftarrow KMeans(H_{P_T}, k_{CH})$
    **for** $i \in [1, k_{CH}]$ **do**
        ▷ Cluster the location of patches in $C_i$
        $C_{i_M} \leftarrow$ KMeans($H_{P_T}(i, :), p_M \times |C_i|$)
        ▷ Construct the Mosaic
        $M \leftarrow C_{i_M}$
    **end for**
    $BoB_I \leftarrow$ Empty array to store BoB index for $I$
    **for** $j \in [1, length(M)]$ **do**
        ▷ Get a patch ($s_h \times s_h$) at $m_x^{idx}$ magnification
        $P_{m_x^{idx}}[j] \leftarrow$ GetPatch($I, M[j]$)
        ▷ Extract the feature from a deep network
        $F \leftarrow$ DeepNet($P_{m_x^{idx}}[j]$)
        ▷ Convert the feature to a barcode
        $B \leftarrow$ MinMaxBarcode($F$)
        Append $B$ to a BoB array $BoB_I$
    **end for**
    Return $BoB_I$
**end procedure**

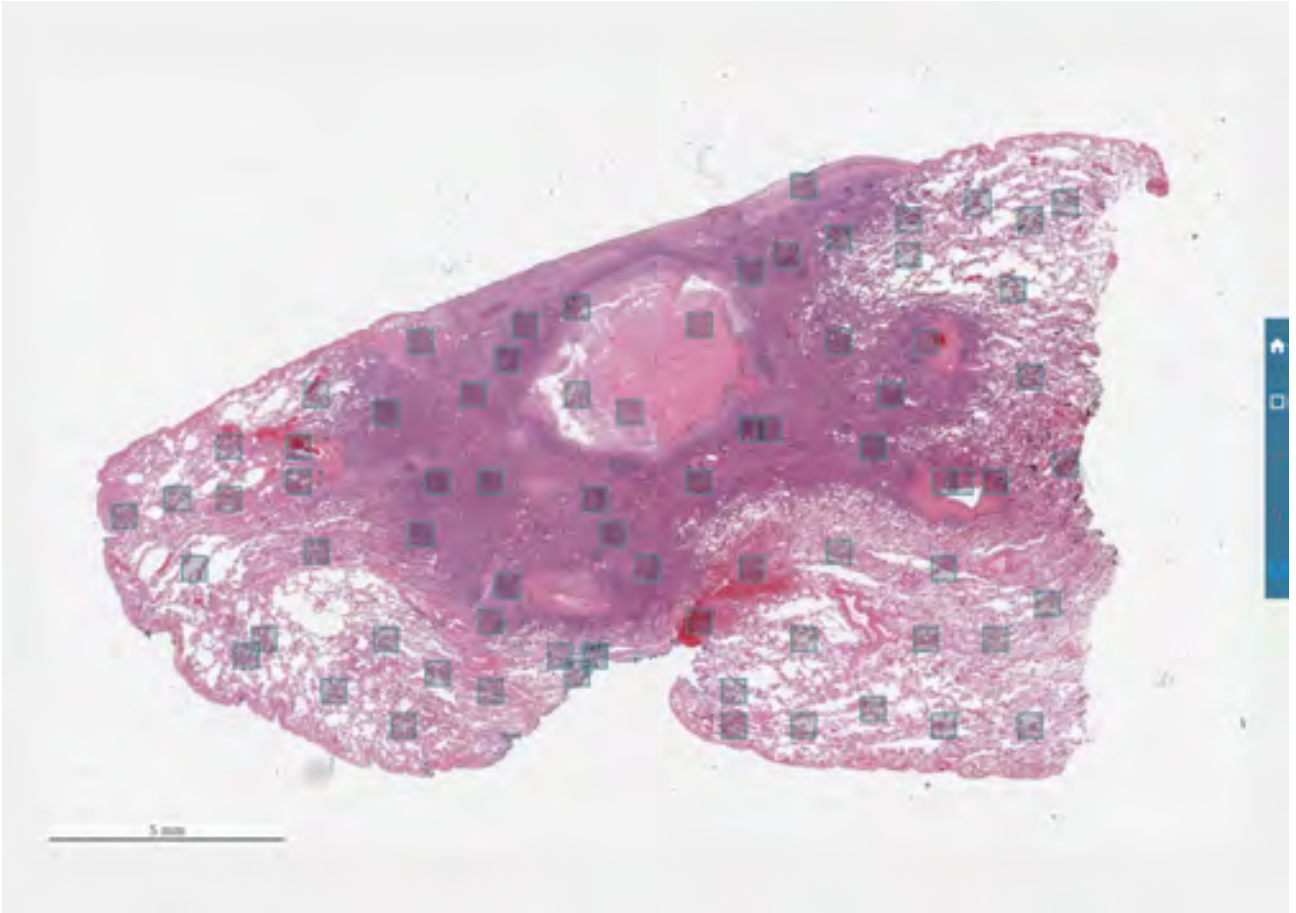

(a) A Lung WSI diagnosed with Wegners Granulomatosis

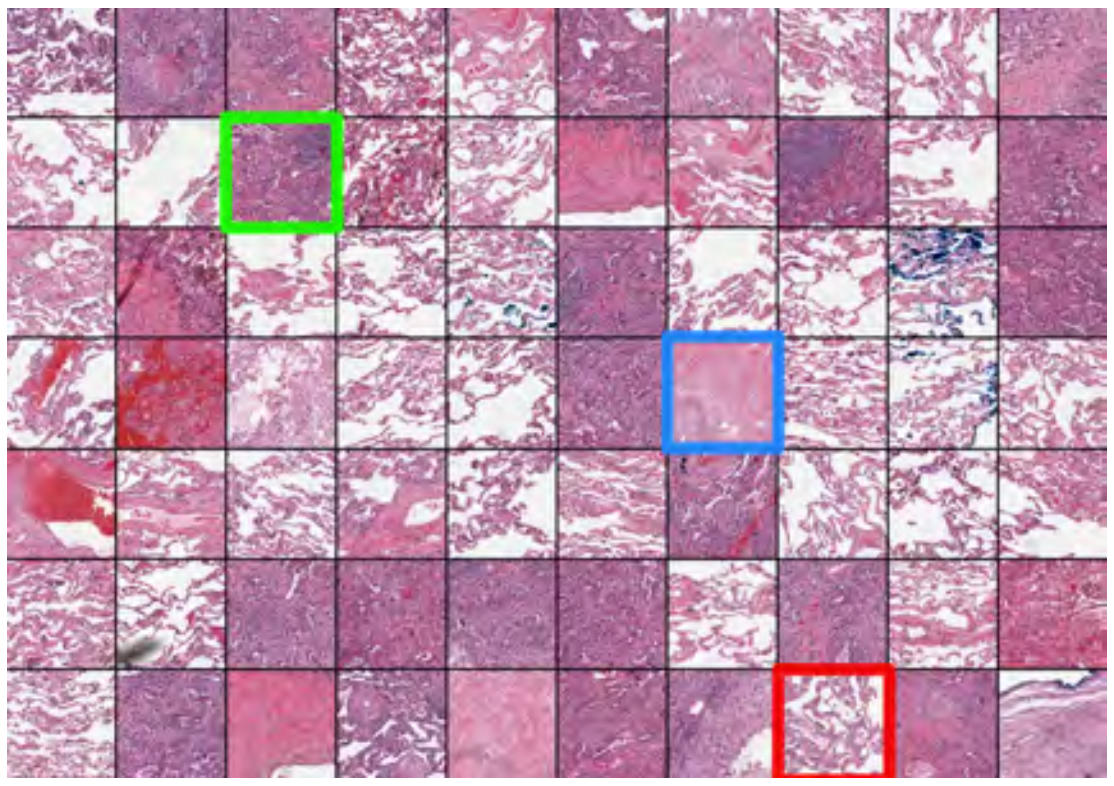

(b) Mosaic consisting of 70 patches highlighted in (a)

?

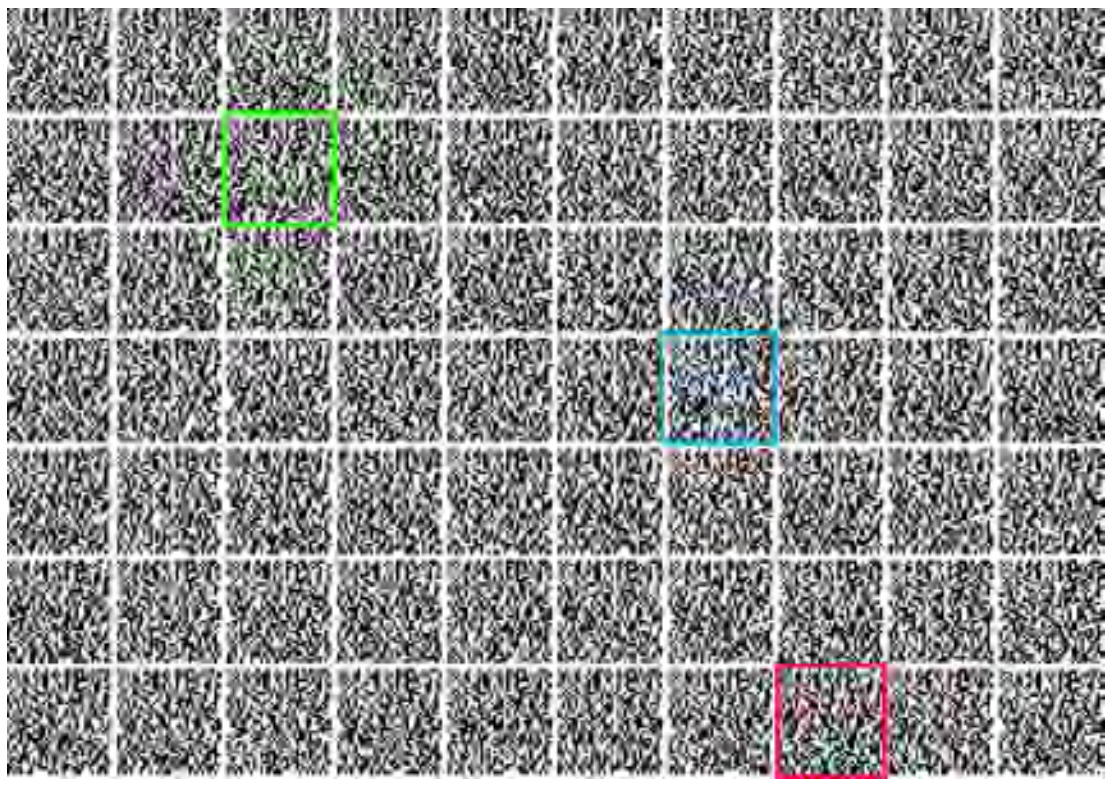

(c) BoB – A bunch of barcodes (binary version of the  mosaic)

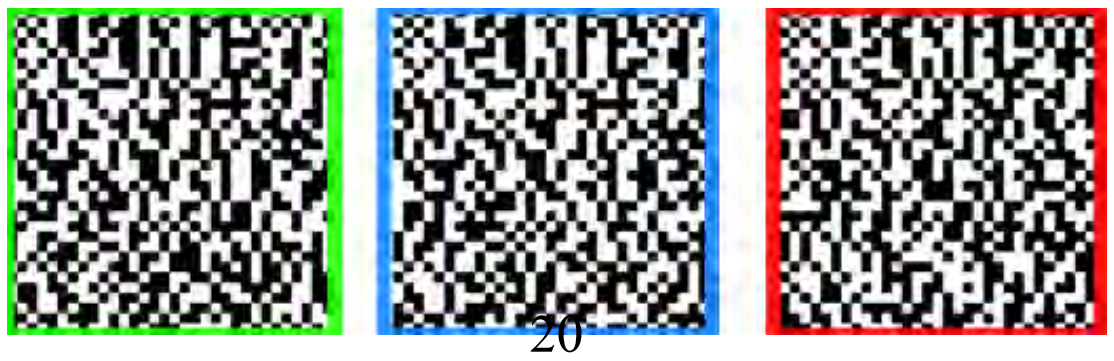

(d) Barcodes of three of the highlighted patches in (b)

Figure 4: Indexing of a sample WSI with 52, 000 × 33, 596 pixels yielding up to 1,200 patches.

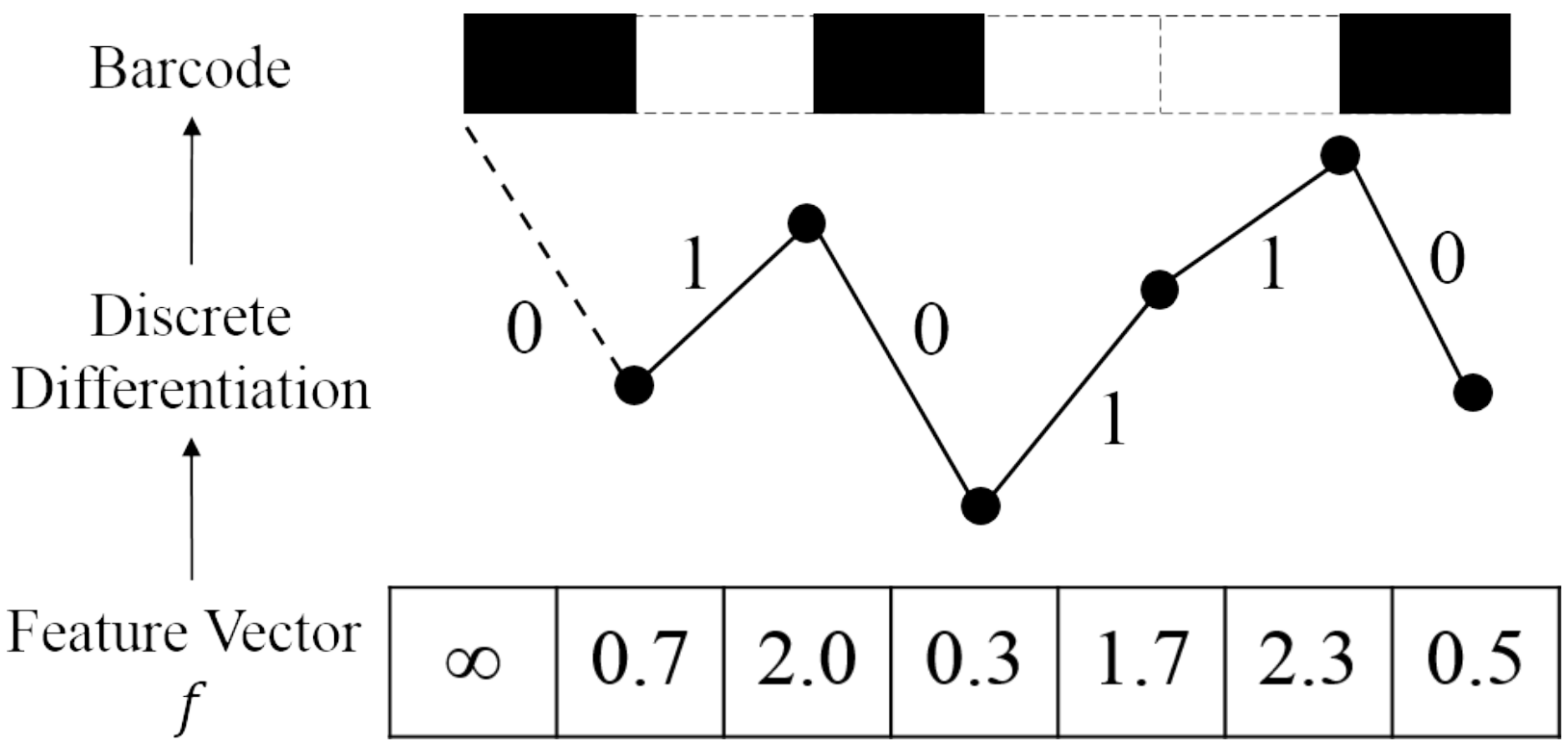

Figure 5: Visual depiction of the *MinMax algorithm* (Tizhoosh et al., 2016) used to convert a feature vector into a barcode for one patch belonging to a specific mosaic.

**Algorithm 2** Distance between two given WSIs $I_q$ and $I$

**procedure** Scan_Distance($I_q$, $I$)
    $D_I \leftarrow \varnothing$
    **for** $b_{I_q} \in I_q.bob$ **do**
        $H_{min} \leftarrow \infty$
        **for** $b_I \in I.bob$ **do**
            $d \leftarrow$ getHammingDistance($b_I, b_{I_q}$)
            **if** $d < H_{min}$ **then**
                $H_{min} \leftarrow d$
            **end if**
        **end for**
        $D_I = D_I \cup \{H_{min}\}$
    **end for**
    $D \leftarrow$ findMedian($D_I$)
    **return** $D$
**end procedure**

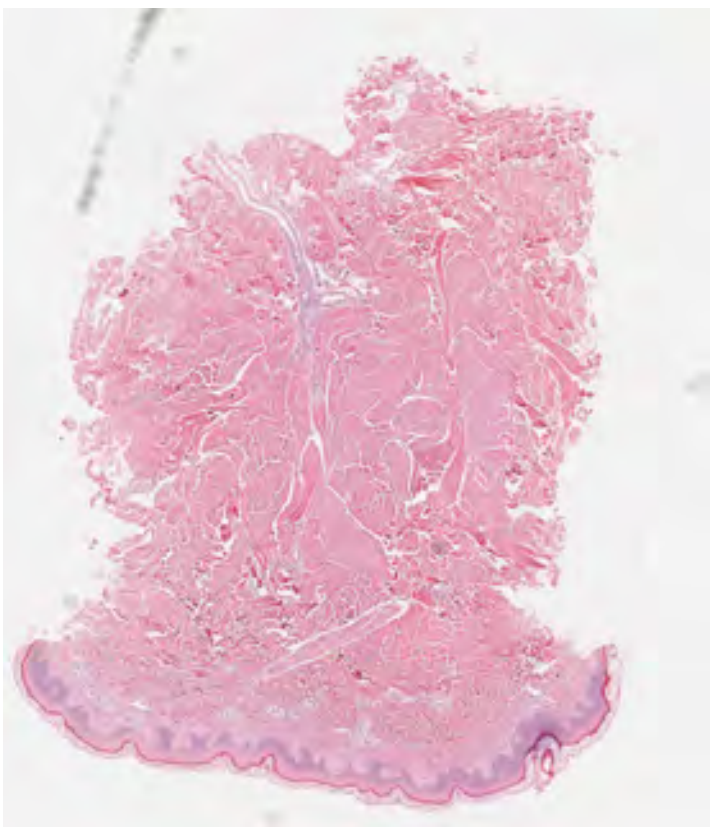

(a) A sample WSI

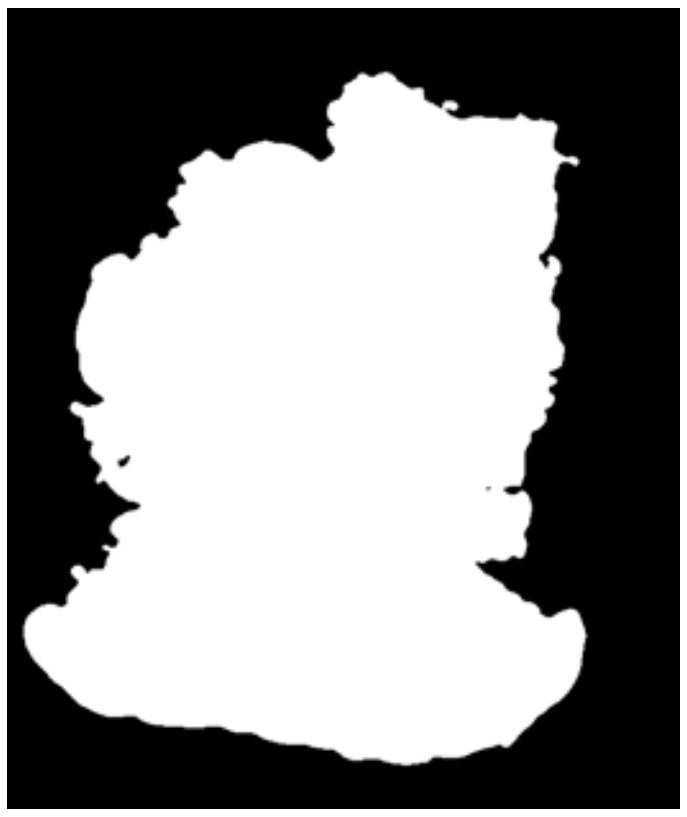

(b) Segmentation mask

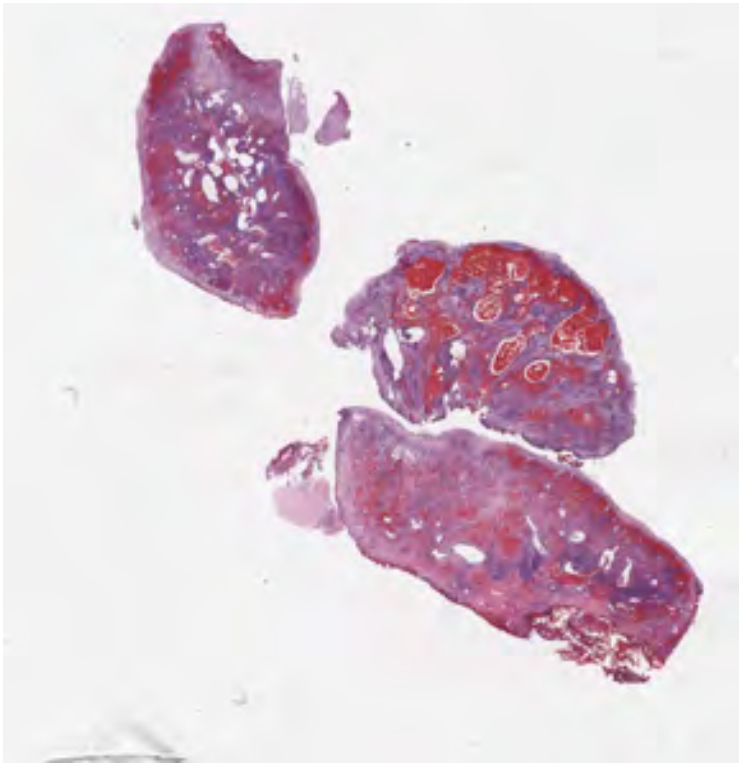

(c) A sample WSI

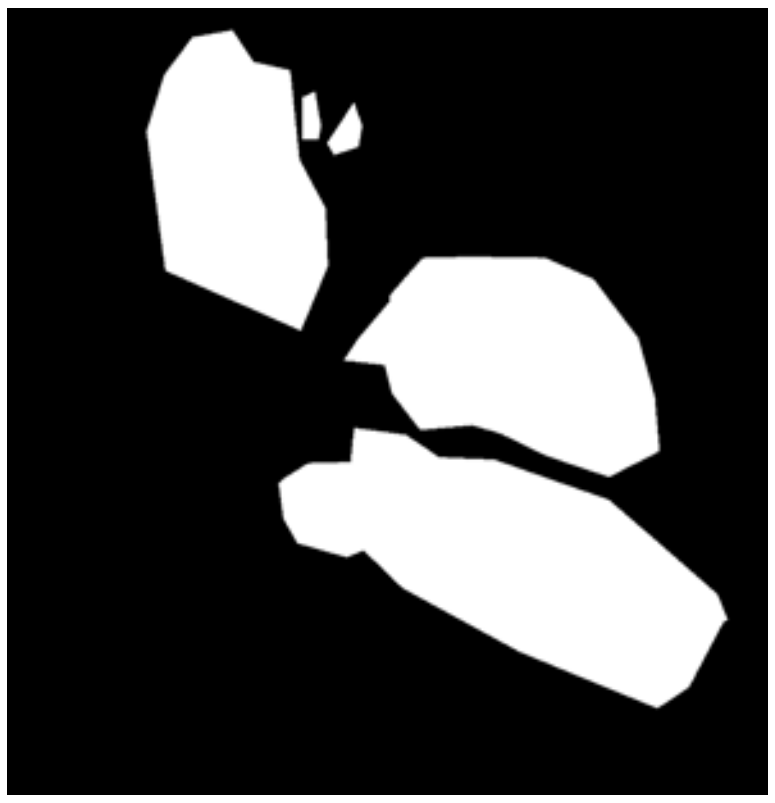

(d) Segmentation mask

Figure 6: (a) WSI from skin sample diagnosed with *collagenoma*, and (b) its tissue segmentation mask, (c) WSI from bladder sample with *high grade urothelial carcinoma*, and (d) its segmentation mask.

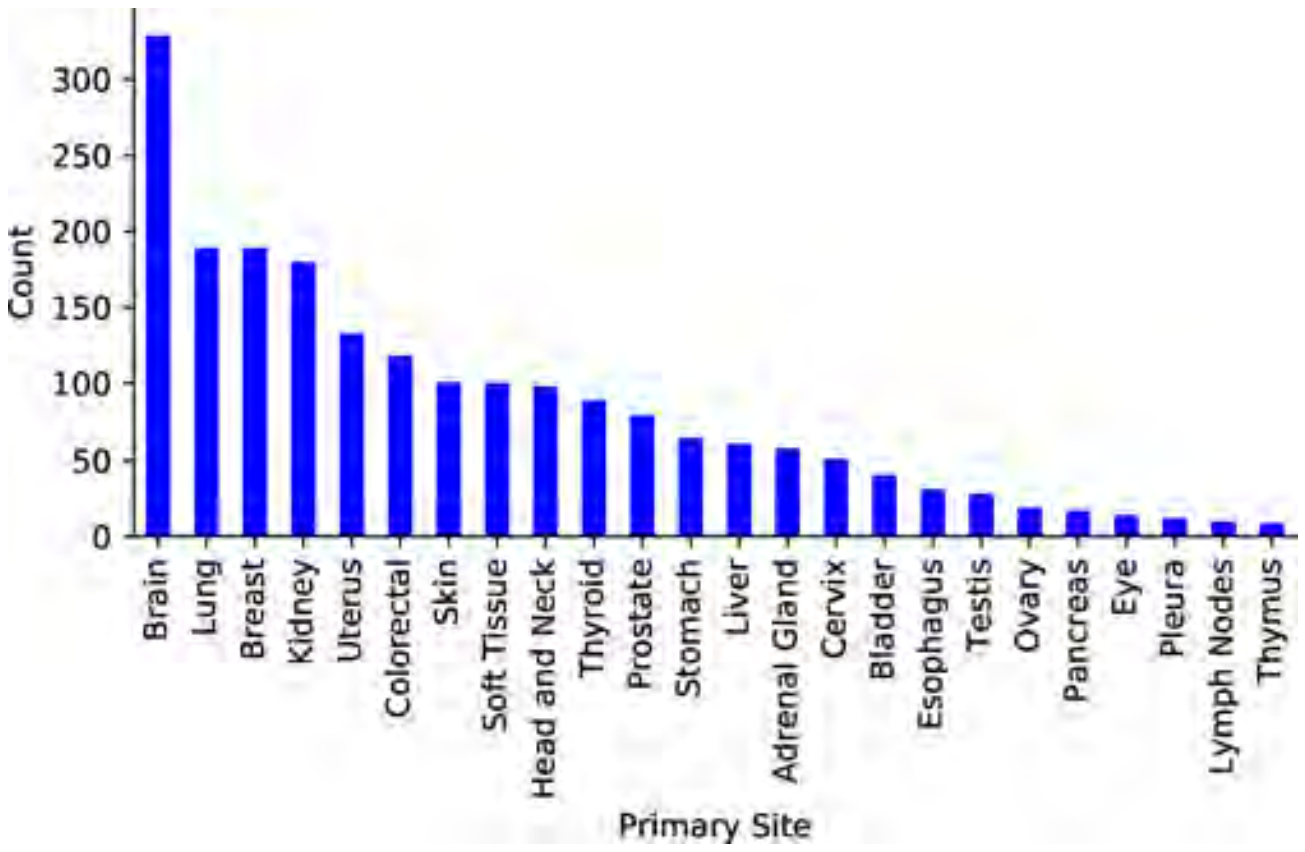

Figure 7: Distribution of WSIs across 24 different primary sites within the NIH dataset (subset of TCGA) containing 2020 WSIs.

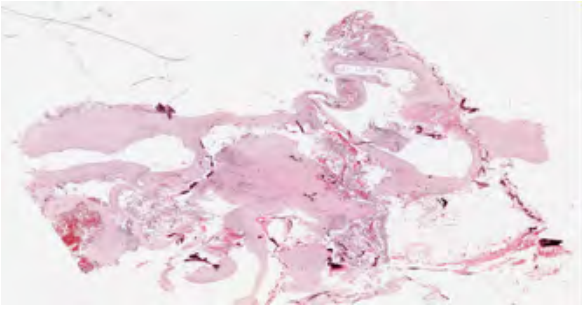

(a) Tibia, Ganglion Cyst, 52,000×38,876

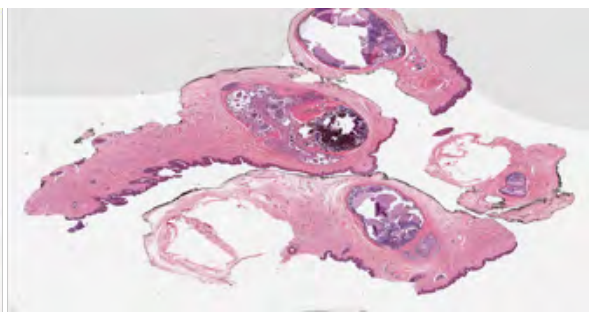

(b) Scrotum, Tumoral Calcinosis, 52,000×39,516

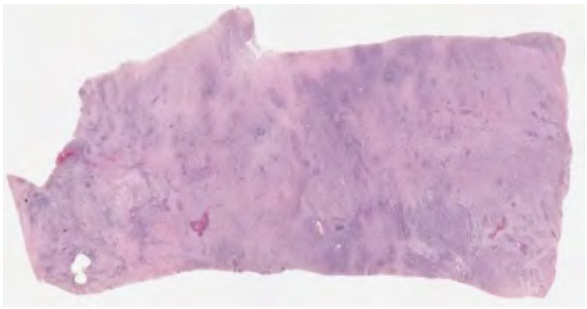

(c) Pelvis, Malignant Peripheral Nerve Sheath Tumor, 48,000×22,920

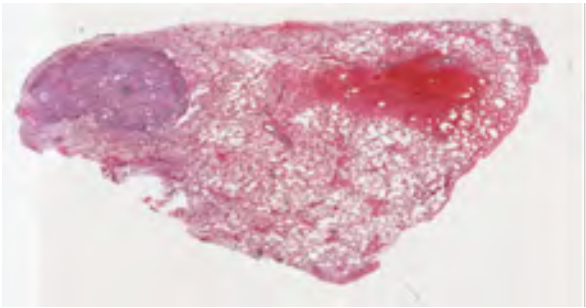

(d) Lung, Metstatic Mature Teratoma, 58,000×45,423

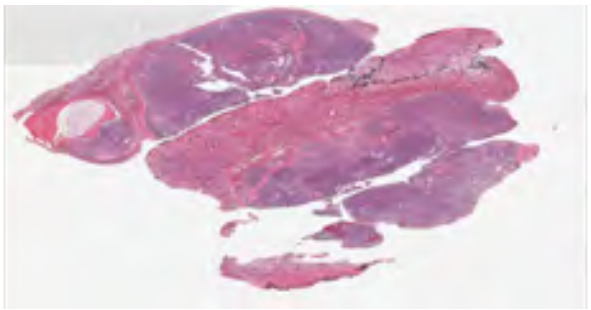

(e) Thyroid, NIFTP, 48,000×45,584

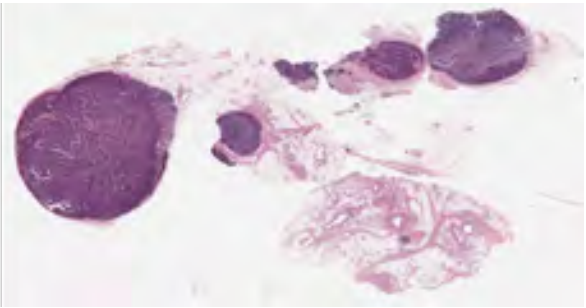

(f) Lymph Node, Metastatic Carcinoid, 48,000×35,665

Figure 8: Six sample WSIs from the UPMC dataset.

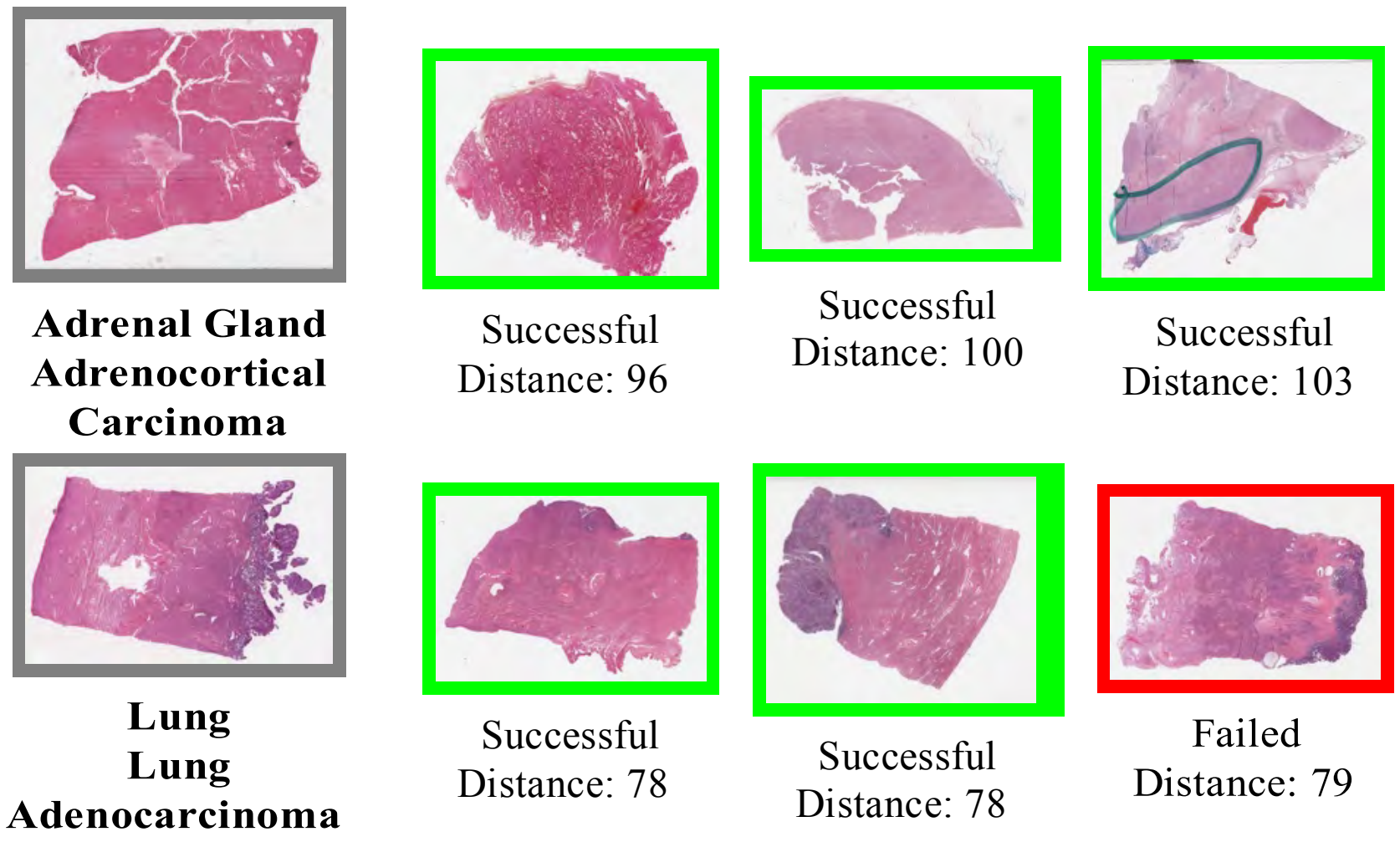

Figure 9: Two sample searches for WSIs from the NIH dataset along with their top similar retrieved images. The failed case is lung squamous cell carcinoma.

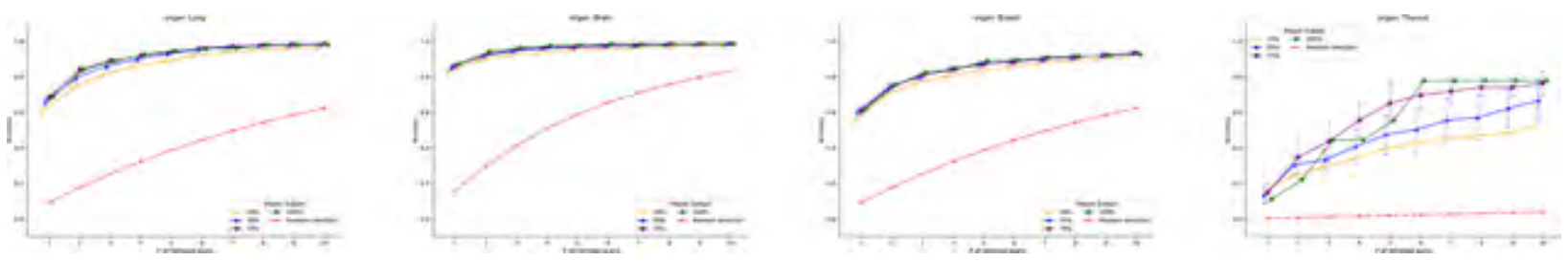

(a) Lung (b) Brain (c) Breast (d) Thymus

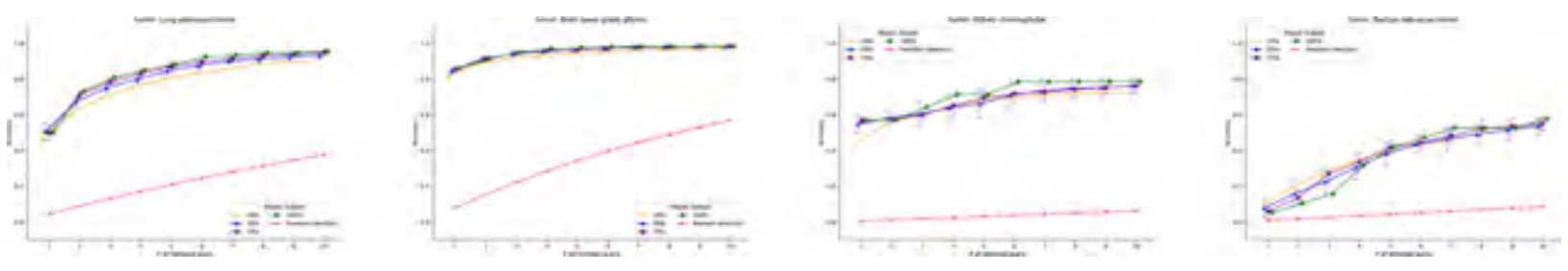

(e) Lung Adenocarcinoma (f) Brain Lower Grade Glioma (g) Kidney Chromophobe (h) Rectum Adenocarcinoma

Figure 10: The accuracy of different retrieval experiments on various primary diagnoses and sites. The red curves show the same accuracy for random retrievals.

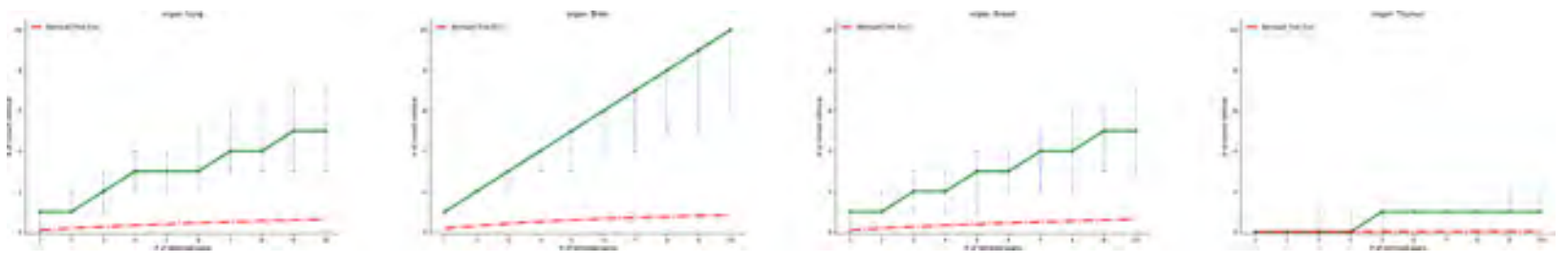

(a) Lung (b) Brain (c) Breast (d) Thymus

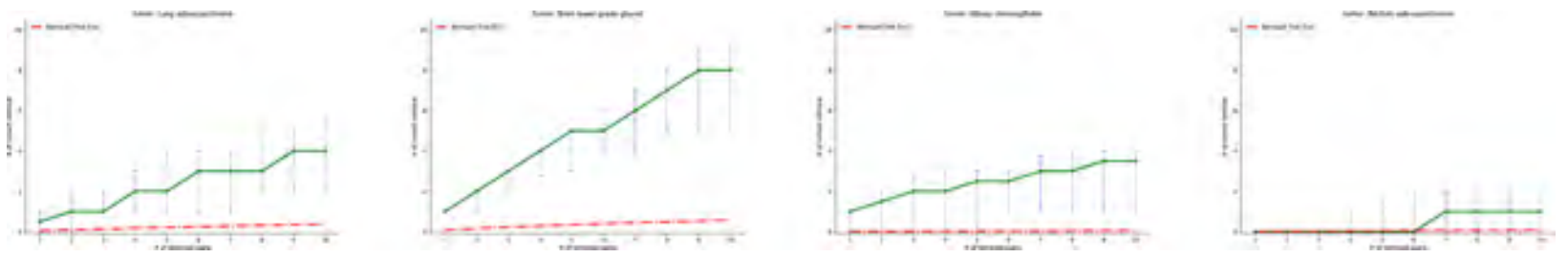

(e) Lung Adenocarcinoma (f) Brain Lower Grade Glioma (g) Kidney Chromophobe (h) Rectum Adenocarcinoma

Figure 11: Median values of the number of correctly retrieved scans for each search experiment on the same primary diagnoses and sites as in Fig. 10.

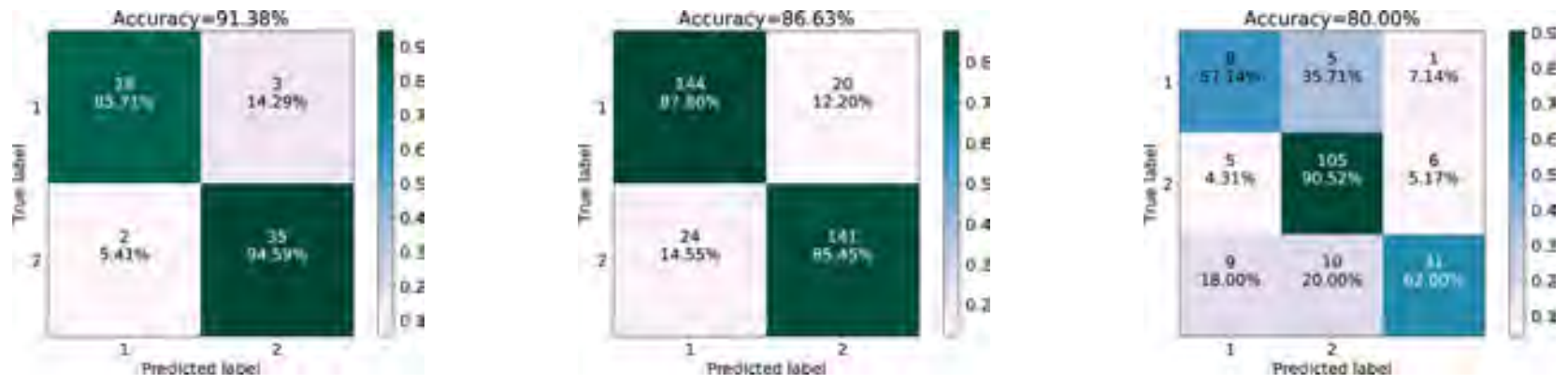

(a) Adrenal Gland (b) Brain (c) Kidney

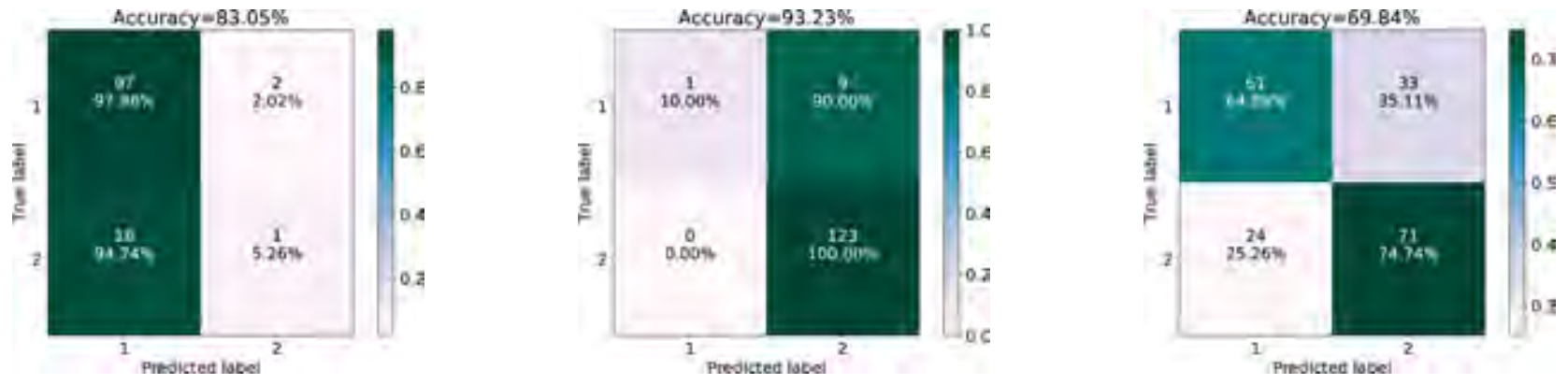

(d) Colorectal (e) Uterus (f) Lung

Figure 12: Confusion matrices for the classification of primary diagnoses for vertical search. The primary diagnosis is assigned by taking a majority vote among the top five search results. Different classes of confusion matrices are (a) 1. Adrenocortical Carcinoma and 2. Pheochromocytoma and Paraganglioma, (b) 1. Brain Lower Grade Glioma and 2. Glioblastoma Multiforme, (c) 1. Kidney Chromophobe, 2. Kidney Renal Clear Cell Carcinoma and 3. Kidney Renal Papillary Cell Carcinoma, (d) 1. Colon Adenocarcinoma and 2. Rectum Adenocarcinoma, (e) 1. Uterine Carcinosarcoma and 2. Uterine Corpus Endometrial Carcinoma, (f) 1. Lung Adenocarcinoma and 2. Lung Squamous Cell Carcinoma.

| | Non Expert | Pathologist Response | | | | | | |
|---|---|---|---|---|---|---|---|---|
| | RC | RC | $Q_1$ | | $Q_2$ | | $Q_3$ | |
| | $\rho_{sp}$ ($\mu \pm \sigma$) | $\rho_{sp}$ | Selection | $d_1$ | Selection | $d_1$ | Selection | $d_1$ |
| 1. | 0.25 ± 0.86 | -0.87 | Bad | 72 | Good | 76 | Good | 77 |
| 2. | 0.87 ± 0.00 | NA | Great | 55 | Great | 58 | Great | 59 |
| 3. | 0.29 ± 1.00 | -0.87 | VeryBad | 68 | VeryBad | 71 | Bad | 72 |
| 4. | -0.31 ± 0.676 | -0.87 | Neutral | 68 | Great | 77 | Great | 80 |
| 5. | -0.16 ± 0.683 | -1.00 | Good | 81 | Great | 86 | Great | 86 |
| 6. | -0.17 ± 0.606 | -0.50 | Neutral | 62 | Great | 63 | Good | 71 |
| 7. | 0.43 ± 0.64 | -0.87 | Good | 73 | Good | 75 | Great | 77 |
| 8. | -0.43 ± 0.612 | NA | Great | 66 | Great | 67 | Great | 69 |
| 9. | -0.22 ± 0.829 | -0.87 | Bad | 76 | Good | 77 | Good | 79 |
| 10. | 0.47 ± 0.77 | 0.87 | Good | 102 | VeryBad | 104 | VeryBad | 105 |
| 11. | 0.43 ± 0.64 | 0.87 | Neutral | 70 | Neutral | 80 | Bad | 81 |
| 12. | -0.58 ± 0.500 | -0.87 | Neutral | 51 | Great | 52 | Great | 55 |
| 13. | NA | NA | Great | 74 | Great | 74 | Great | 74 |
| 14. | -0.87 ± 0.000 | NA | Great | 58 | Great | 65 | Great | 69 |
| 15. | -0.55 ± 0.356 | -0.50 | VeryBad | 77 | VeryBad | 78 | Good | 78 |
| 16. | 0.98 ± 0.05 | 1.0 | Great | 87 | Good | 93 | VeryBad | 97 |
| 17. | -0.50 ± - | NA | Great | 62 | Great | 62 | Great | 64 |
| 18. | -0.50 ± 0.707 | NA | Great | 91 | Great | 95 | Great | 95 |
| 19. | -0.58 ± 0.500 | -0.87 | Good | 68 | Great | 73 | Great | 75 |
| 20. | -0.07 ± 0.493 | NA | Neutral | 71 | Neutral | 76 | Neutral | 79 |
| 21. | 0.07 ± 0.49 | 0.50 | Great | 70 | Neutral | 71 | Good | 78 |
| 22. | 0.91 ± 0.06 | 0.87 | Great | 58 | Bad | 74 | Bad | 77 |
| 23. | 0.42 ± 0.49 | 0.00 | Good | 79 | Great | 83 | Neutral | 83 |
| 24. | 0.87 ± 0.00 | 0.8 | Great | 57 | Great | 60 | VeryBad | 77 |
| 25. | -0.12 ± 0.694 | 0.0 | Great | 66 | Good | 78 | Great | 79 |
| 26. | -0.77 ± 0.255 | -0.87 | Neutral | 55 | Good | 55 | Great | 60 |
| 27. | 0.58 ± 0.70 | 1.00 | Good | 72 | Neutral | 73 | Bad | 75 |
| 28. | 0.93 ± 0.07 | 1.00 | Great | 77 | Good | 79 | Bad | 82 |
| 29. | 0.52 ± 0.77 | 0.5 | Great | 50 | Neutral | 70 | Good | 74 |
| 30. | 0.59 ± 0.18 | -0.50 | Good | 75 | Good | 76 | Great | 76 |
| 31. | 0.65 ± 0.43 | 0.87 | Great | 65 | Great | 70 | Good | 76 |
| 32. | 0.65 ± 0.40 | NA | Great | 85 | Great | 90 | Great | 91 |
| 33. | 0.87 ± 0.00 | NA | Great | 58 | Great | 62 | Great | 65 |
| 34. | -0.17 ± 0.408 | -0.87 | Bad | 76 | Bad | 81 | Great | 83 |
| 35. | 0.87 ± NA | NA | Great | 75 | Great | 78 | Great | 79 |
| 36. | -0.62 ± 0.750 | NA | Great | 52 | Great | 56 | Great | 56 |
| 37. | NA | NA | Great | 60 | Great | 61 | Great | 62 |
| 38. | 0.91 ± 0.06 | 1.00 | Great | 71 | Bad | 76 | VeryBad | 83 |
| 39. | 0.87 ± 0.00 | NA | Great | 65 | Great | 69 | Great | 71 |
| 40. | -0.25 ± 0.866 | NA | Good | 67 | Good | 68 | Good | 68 |
| 41. | 0.25 ± 0.61 | 0.50 | Neutral | 77 | Good | 77 | Neutral | 82 |
| 42. | 0.42 ± 0.20 | 0.87 | Good | 89 | Neutral | 96 | Bad | 96 |
| 43. | 0.17 ± 0.38 | 0.0 | Great | 58 | Good | 63 | Great | 68 |
| 44. | -0.14 ± 0.725 | NA | Great | 83 | Great | 94 | Great | 100 |
| 45. | 0.87 ± NA | 0.87 | Great | 54 | Good | 60 | Good | 62 |
| 46. | 0.13 ± 0.59 | 0.87 | Neutral | 74 | Bad | 75 | Bad | 76 |
| 47. | -0.17 ± 0.764 | NA | Great | 57 | Great | 60 | Great | 60 |
| 48 | -0.27 ± 0.729 | -0.87 | Bad | 60 | Neutral | 63 | Neutral | 68 |

Table 3: Response of pathologist for each question asked during the experiment series (subsection 5.3). Shades of green represent the positive responses (in favour of Yottixel) and shades of red represent negative responses (against Yottixel). Rank coefficient (RC) represents the rank correlation of the response with respect to the internal ranking of Yottixel based on the Hamming distance. Best viewed coloured.

| Reviewer | Ranking Coeff. | Selected Option | | | | | | | | | | | |
|---|---|---|---|---|---|---|---|---|---|---|---|---|---|
| | | Disagreement: `Bad` or `Very Bad` | | | | Neutral | | | | Agreement: `Good` or `Great` | | | |
| | $\rho_{sp}$ | $d_{\ell_1} (\mu \pm \sigma)$ | $p(Q_1)$ | $p(Q_2)$ | $p(Q_3)$ | $d_{\ell_1} (\mu \pm \sigma)$ | $p(Q_1)$ | $p(Q_2)$ | $p(Q_3)$ | $d_{\ell_1} (\mu \pm \sigma)$ | $p(Q_1)$ | $p(Q_2)$ | $p(Q_3)$ |
| **Pathologist** | 0.040 ± 0.802 | 79.541 ± 10.814 | 0.250 | 0.291 | 0.458 | 72.052 ± 10.298 | 0.421 | 0.368 | 0.210 | 71.366 ± 11.674 | 0.336 | 0.336 | 0.326 |
| Person 1 | 0.049 ± 0.713 | 78.761 ± 11.835 | 0.261 | 0.404 | 0.333 | 76.555 ± 10.240 | 0.333 | 0.296 | 0.370 | 68.146 ± 10.113 | 0.373 | 0.306 | 0.320 |
| Person 2 | 0.211 ± 0.724 | 80.047 ± 10.283 | 0.238 | 0.238 | 0.523 | 77.064 ± 11.048 | 0.322 | 0.322 | 0.354 | 69.739 ± 11.109 | 0.358 | 0.358 | 0.282 |
| Person 3 | 0.206 ± 0.669 | 71.263 ± 13.341 | 0.210 | 0.421 | 0.368 | 76.689 ± 11.747 | 0.241 | 0.310 | 0.448 | 71.958 ± 11.200 | 0.385 | 0.322 | 0.291 |
| Person 4 | 0.038 ± 0.745 | 76.062 ± 13.273 | 0.250 | 0.312 | 0.437 | 75.131 ± 11.135 | 0.342 | 0.394 | 0.263 | 71.266 ± 11.474 | 0.344 | 0.311 | 0.344 |
| Person 5 | 0.315 ± 0.704 | 79.176 ± 10.387 | 0.058 | 0.352 | 0.588 | 75.194 ± 10.400 | 0.250 | 0.416 | 0.333 | 70.692 ± 11.889 | 0.417 | 0.296 | 0.285 |
| Person 6 | 0.272 ± 0.657 | 81.500 ± 12.036 | 0.071 | 0.357 | 0.571 | 74.928 ± 8.785 | 0.285 | 0.285 | 0.428 | 71.049 ± 11.803 | 0.382 | 0.343 | 0.274 |

Table 4: Proportion of participant's selected option for each $Q_1$, $Q_2$, $Q_3$ across three broad categories: Disagreement, Neutral, and Agreement. The average Hamming distances $d_{\ell_1}$ for each category shows the decreasing trend from $Q_1$ to $Q_3$. The positive ranking coefficient $\rho_{sp}$ corresponds to the positive correlation of the ordering of the participant with respect to the search results.

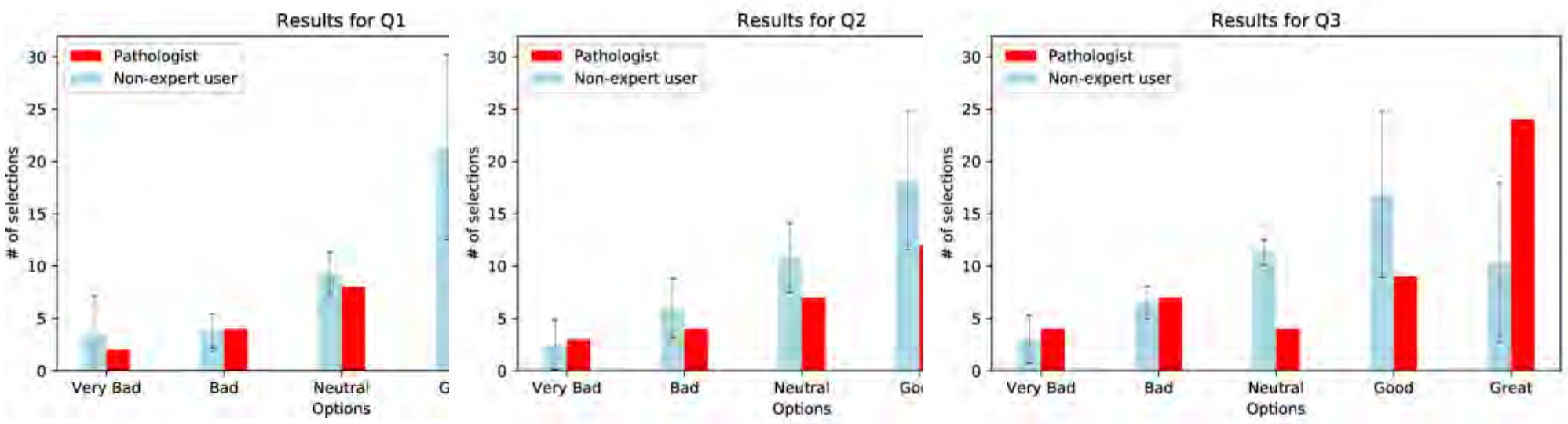

(a) Response frequency for $Q_1$. (b) Response frequency for $Q_2$ (c) Reponse frequency for $Q_3$

Figure 13: Response frequency for each option among the top three search results. There are more selections of `Bad` and `Very Bad` for $Q_3$ compared with $Q_1$ and $Q_2$.

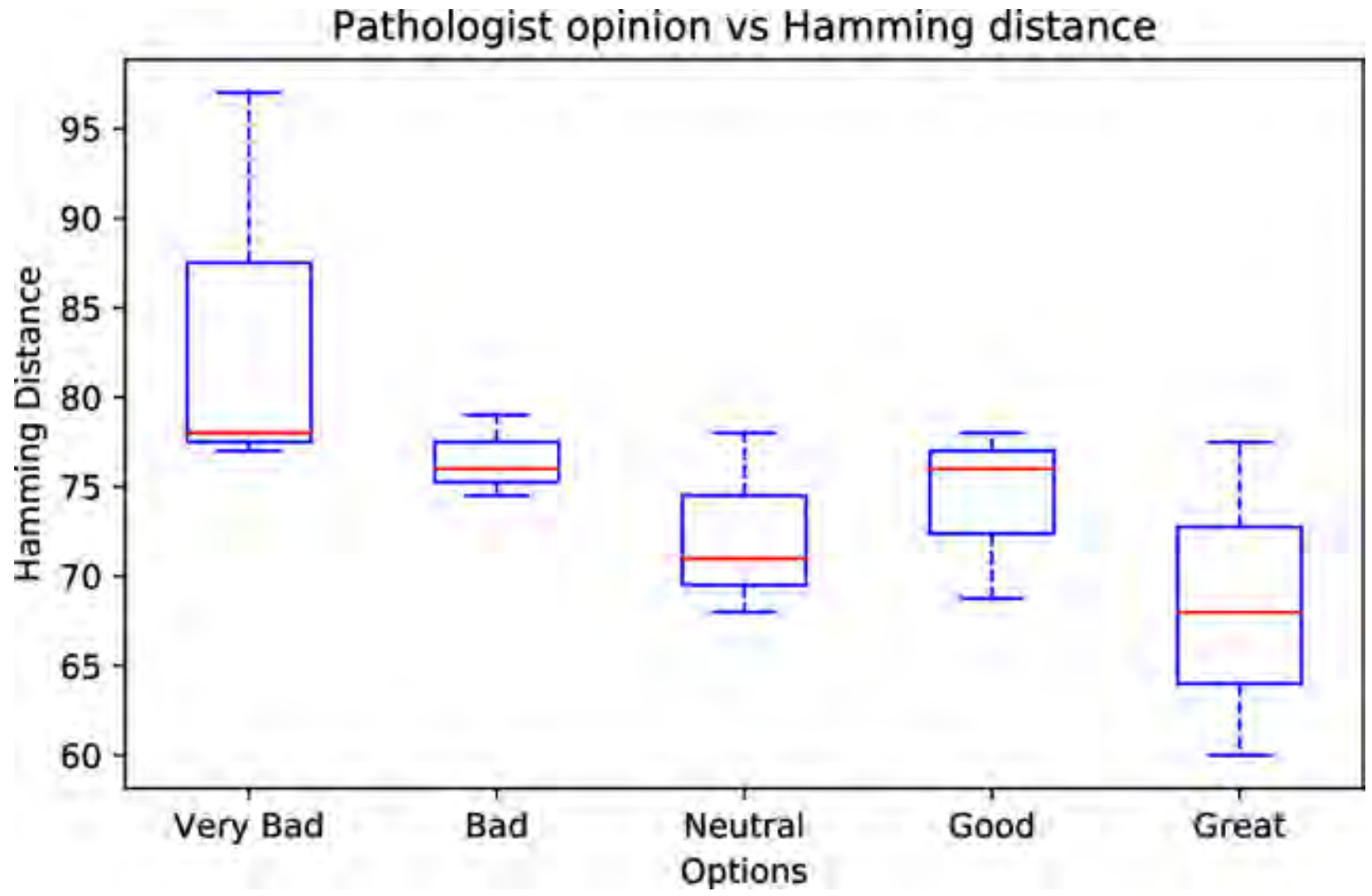

Figure 14: Negative correlation between the Yottixel and the pathologist's selection is evident through the box-plot. The `Great` selection has the least median Hamming distance.